\documentclass[a4paper,11pt]{article}
\usepackage{jinstpub} 
\usepackage{lineno}
\usepackage{upgreek,xspace}
\usepackage{subfigure}
\usepackage{caption}
\usepackage{subcaption}
\def\textmu{\ensuremath\upmu}

\title{Development of wavelength-shifting PEN foils for next generation experiments}

\author[a,1]{M.~Kuźniak\note{Corresponding author.}}
\author[b]{S.~Pawłowski}
\author[b]{A.~Abramowicz}
\author[a]{A.\,F.\,V.~Cortez}
\author[b]{M.~Kumosiński}
\author[c]{T.~\L ęcki}
\author[a]{G.~Nieradka}

\affiliation[a]{AstroCeNT, Nicolaus Copernicus Astronomical Center of the Polish Academy of Sciences, Rektorska 4, 00-614 Warsaw, Poland}
\affiliation[b]{Łukasiewicz Research Network – Industrial Chemistry Institute, Rydygiera 8, 01-793 Warsaw, Poland}
\affiliation[c]{Biological and Chemical Research Centre, Faculty of Chemistry, University of Warsaw, \.Zwirki i Wigury 101, 02-089 Warsaw, Poland}

\emailAdd{mkuzniak@camk.edu.pl}

\abstract{Polyethylene naphthalate (PEN) foils have been demonstrated as a wavelength shifter suitable for operation in liquid argon. At the same time, wavelength shifting efficiency of technical grades of PEN, commercially available on the market, is lower than that of tetraphenyl butadiene (TPB). This paper reports on an R\&D program focused on exploring the intrinsic limitations of PEN and optimizing it for the highest achievable wavelength shifting efficiency.
}

\keywords{wavelength shifters, polyethylene naphthalate, liquid argon detectors}

\begin{document}
\maketitle
\flushbottom

\section{Introduction}
Liquid argon detectors for dark matter searches, long-baseline neutrino experiments and neutrinoless double-beta decay searches, among other applications, rely on detection of vacuum ultraviolet (VUV) argon scintillation light at 128~nm. Wavelength shifters are commonly used to convert it to visible blue light, where the photon detection efficiency of available photosensors is significantly higher~\cite{instruments5010004}. 

Polyethylene naphthalate (PEN), available as large format polymeric film, has been proposed~\cite{Kuźniak2019} as a scalable and stable alternative to traditionally employed tetraphenyl butadiene (TPB). TPB coatings are deposited via vacuum evaporation and hence challenging to scale up to 100-10000~m$^2$ of surface area, anticipated in the next generation detectors.

Commercially available, technical, PEN foil grades, however, which are neither made nor optimized for this application, provide the wavelength shifting efficiency (WLSE) of approx. 34-70\% of TPB~\cite{Boulay2021, Abraham_2021, Araujo2022}. Maximizing WLSE is essential for dark matter detectors, as it lowers their energy threshold. Due to the steeply falling WIMP-induced nuclear recoil spectrum this translates into a non-linear increase in sensitivity. At the same time, the literature suggests that better PEN performance can be achieved~\cite{Kuźniak2019}. Potential factors negatively affecting the WLSE of commercial PEN films include: 
\begin{itemize}
    \item over-coatings or additives (e.g. fillers, colorants, plasticizers, stabilizers, catalysts), which improve the mechanical properties of the foil in industrial applications, while degrading its optical yield,
    \item unknown history of samples (storage, integrated UV exposure),
    \item crystalline structure and orientation (e.g. amorphous, bi-axially stretched),
    \item thermal history (sequence and details of cooling/warming cycles experienced by the sample since the last melted state, e.g. during production, processing and storage).
\end{itemize}
Recent progress on R\&D aimed at identifying the key factors driving the WLSE of PEN and, informed by that, synthesising a grade optimized for application in liquid argon detectors, is reported here. 

\section{Analysis of commercial grades}
\begin{figure}[t]
    \centering
    \subfigure[1$^\mathrm{st}$ heating]{\centering\includegraphics[width=0.5\linewidth]{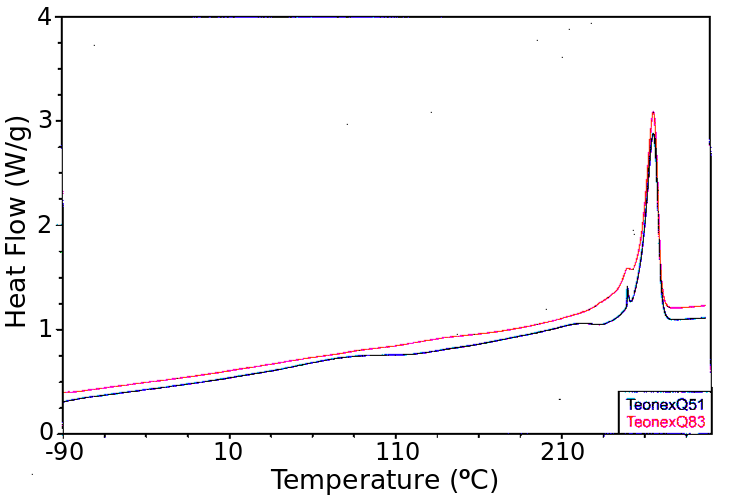}}\subfigure[Cooldown]{\includegraphics[width=0.5\linewidth]{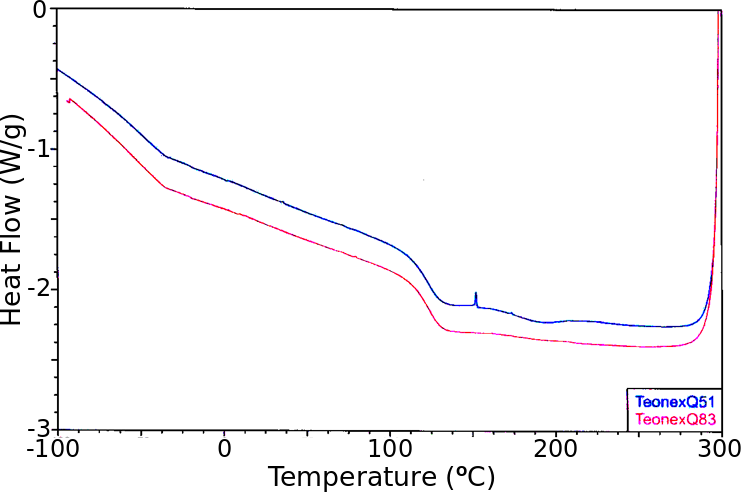}\label{a}}
    \subfigure[2$^\mathrm{nd}$ heating]{\includegraphics[width=0.5\linewidth]{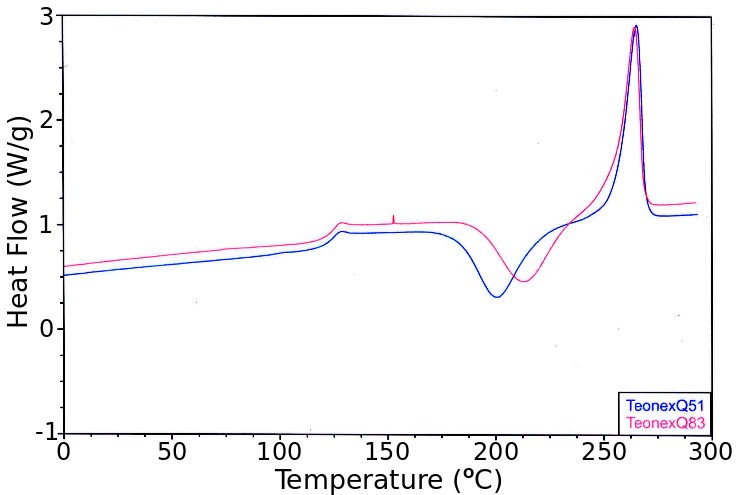}}\subfigure[Teonex Q51 cooldown]{\includegraphics[width=0.50\linewidth]{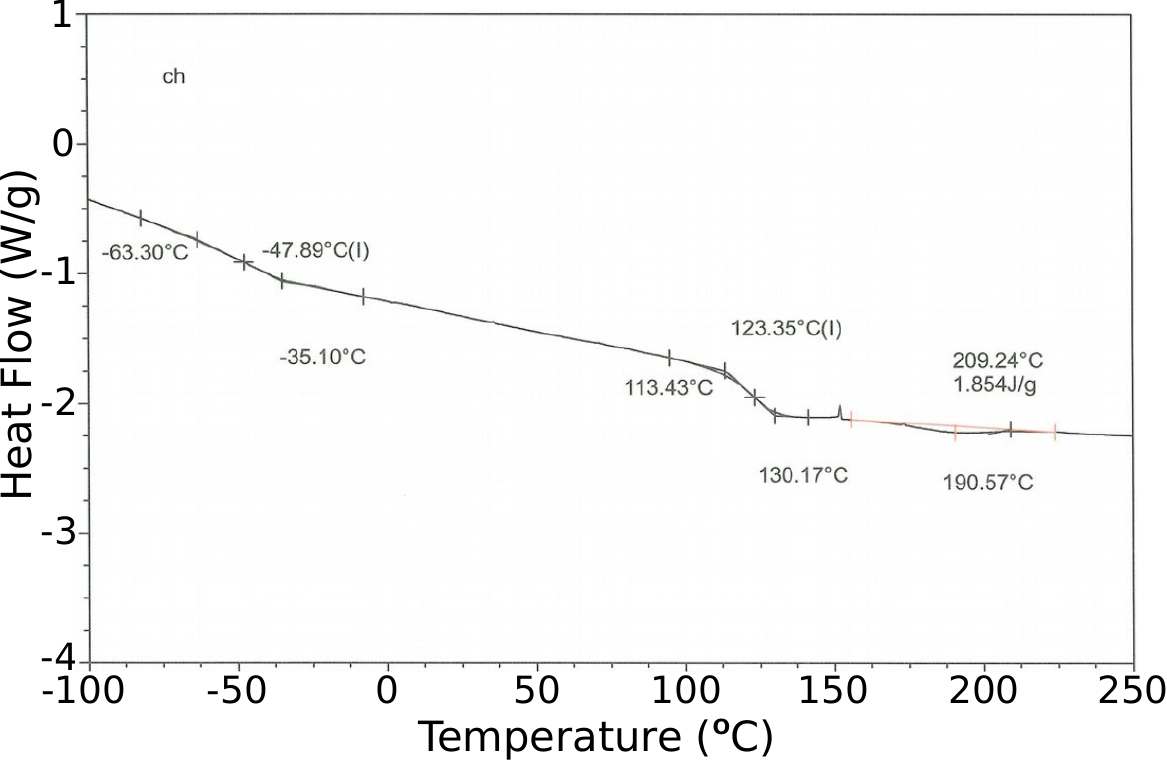}}
    \caption{Differential scanning calorimetry curves for two commercially available PEN grades: Teonex~Q51 and Teonex~Q83. Both samples went through a cycle: (a)~1$^\mathrm{st}$ heating, (b)~cooldown, (c)~2$^\mathrm{nd}$ heating; with (d)~showing zoomed-in cooldown curve of Teonex~Q51 to illustrate the extraction of transition temperatures. Step-like features, minima and maxima correspond to, respectively, glass transitions, crystallization and melting.
    Nearly identical melting temperatures during both heating phases indicate that both samples are made of the same polymer.}
    \label{fig:dsc}
\end{figure}

One of the first steps was an attempt to identify the dominant factors affecting WLSE by examining grades which were the best (Teonex Q51) and worst (Teonex Q83) performing in terms of WLSE, based on the earlier measurements. Differential scanning calorimetry (DSC) is a thermoanalytical technique in which the difference in the amount of heat required to increase the temperature of a sample and reference is measured as a function of temperature. This allows, in particular, to map temperatures of transitions between amorphous and crystalline phases. 
Both samples went through a cycle in the $-100$-300$^\circ$C range, see  Fig.~\ref{fig:dsc}: 1$^\mathrm{st}$ heating, cooldown, and 2$^\mathrm{nd}$ heating. The first melting erases the thermal history of the sample, allowing to compare the intrinsic differences between materials in the subsequent phases of the cycle.
\begin{table}[h]
\centering
 \begin{tabular}{c||r|r|r} 
 Sample & 1$^\mathrm{st}$ heating & Cooldown & 2$^\mathrm{nd}$ heating \\ [0.5ex] 
 \hline
 Teonex Q51 & $T_g$=51, $T_m$=218, 250-266 & $T_c$=191, $T_g$=123, $-48$ & $T_g$=125, $T_c$=201, $T_m$=265 \\ 
 Teonex Q83 & $T_m$=249-266 & $T_g$=124, $-56$ & $T_g$=126, $T_c$=214, $T_m$=265 \\ [1ex] 
 \end{tabular}
 \caption{Summary of temperatures (in $^\circ$C) at which DSC revealed phase transitions in the Teonex samples: cold crystallization ($T_c$), glass transition ($T_g$) and melting ($T_m$).}
\label{tab}
\end{table}

Phase transitions observed with DSC are listed in~Tab.~\ref{tab}. Nearly identical melting temperatures during both heating phases, despite somewhat broader maximum in Teonex~Q83, indicate that the same polymer and of similar molecular mass is present in both samples. Significant differences in glass transition (during cooldown) and cold crystallization temperatures (cooldown and 2$^\mathrm{nd}$ heating) typically indicate either differences in the level of cross-linking (consequence of different molecular masses and purity grades of the base chemicals used for production, including different co-monomers) or difference in the type, or the very presence of, additives.

A nuclear magnetic resonance~(NMR) scan was initially intended to cross-check the DSC results and provide new insights on the nature of impurities or additives. At the end it was not performed due to difficulties in sample preparation required for NMR, i.e. dissolution or fine pulverization of PEN.

\section{UV aging}
Effects of UV aging on fluorescence intensity of PEN have already been reported in the literature~\cite{963512}. Aging test of the Teonex Q51 grade was performed following the PN-EN 60794-1-22 standard, in which the samples were exposed to a Feutron SOL500 lamp, with 650 W/m$^2$ in the 300 to 780~nm spectral range and at elevated temperature inside of a climatic chamber.
\begin{figure}
    \centering
\includegraphics[width=0.38\linewidth]{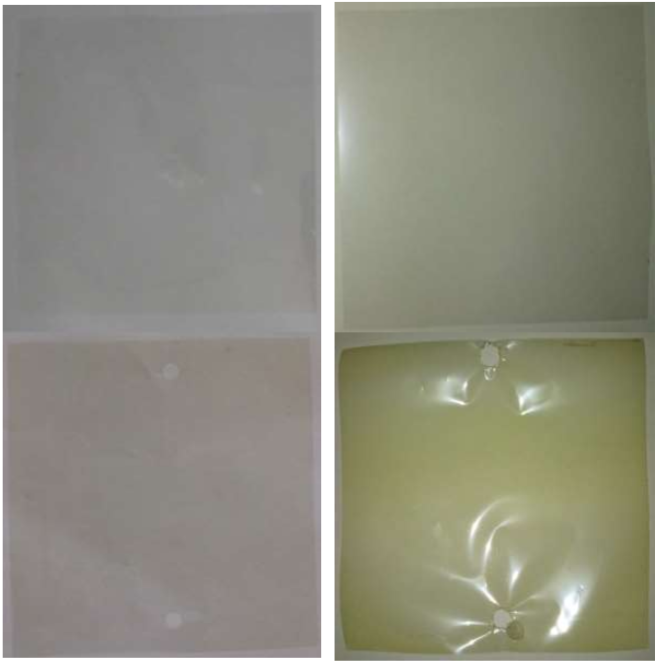}\includegraphics[width=0.6\linewidth]{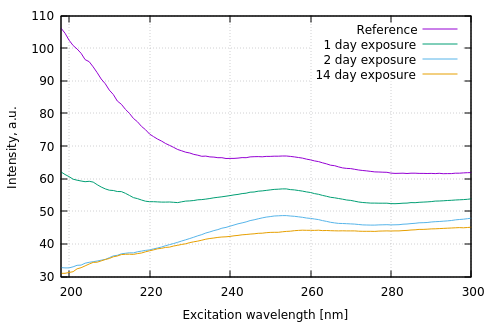}
    \caption{Reference PEN (Teonex Q51) sample, not exposed to UV and kept in the dark (top), next to the sample aged for 1 day (bottom left), and next to a sample after 14 days of exposure to the solar simulator (middle). (Right) Fluorescence emission intensity as a function of excitation wavelength for samples after the respective exposures.}
    \label{fig:uv}
\end{figure}
Significant yellowing and a drop in emission intensity is observed already after 1 day of exposure in the chamber (equivalent to about 2 months of exposure to sunlight), see Fig.~\ref{fig:uv}. The fluorescence intensity has been characterised using a Shimadzu UV-3600 spectrophotometer equipped with a 15~cm diameter integrating sphere accessory (LISR-3100), as in Refs.~\cite{Kuźniak2019,Boulay2021}.

\section{Custom synthesis}
Given the indication from DSC on differences in cross-linking levels or additives between grades, an attempt has been made to synthesize an additive-free batch of PEN, following the process outlined in the original patent~\cite{patent} and using a small glass reactor. Due to limited amount of the product, 100~\textmu m-thick foil samples for fluorescence intensity characterization were made by pressing the material at 275$^\circ$C (typically at least several kg of material are required by foil drawing/stretching machines).

\begin{figure}
    \centering
    \includegraphics[width=0.2\linewidth]{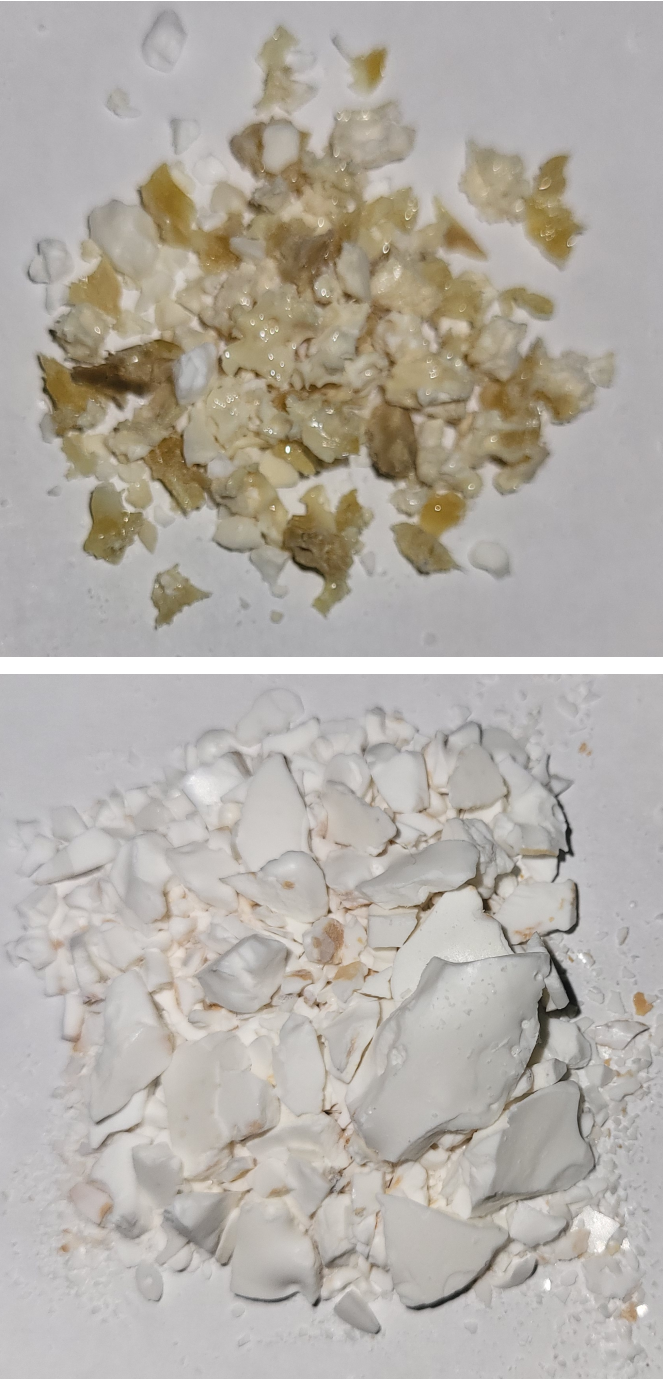}
        \includegraphics[width=0.65\linewidth]{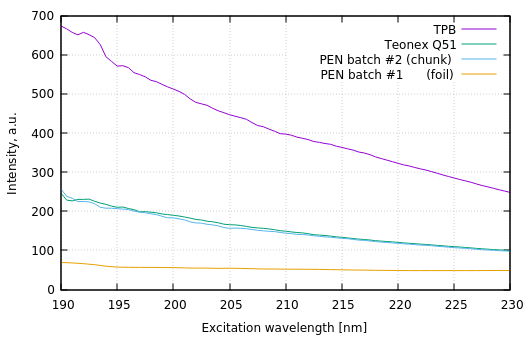}
    \caption{Custom synthesized additive-free PEN. (Top Left) Batch \#1, obtained early in the development process, with signs of yellowing of the product. (Bottom Left) Optimized batch \#2 after fine-tuning the parameters of the process and purification of the base chemicals. (Right) Fluorescence intensity of custom made PEN (batches \#1 and \#2) compared to a commercial grade Teonex~Q51 and a TPB reference, as a function of excitation wavelength, and at room temperature.}
    \label{fig:pen}
\end{figure}

Initial batches suffered from thermal degradation and yellowing of the product caused by
challenges in maintaining uniformly high and hot-spot free temperature distribution throughout the
mixture, as well as impurities in the base chemicals. Additional purification step and better temperature control resulted in a significantly improved white product (batch \#2), see Fig.~\ref{fig:pen}.

The fluorescence intensity of batch~\#2, in the raw form of chunks (as extracted from the reactor), matches that of commercial Teonex Q51 foils. Further intensity increase due to reduced self-absorption is expected after producing a thin foil out of batch \#2 material, potentially marginally surpassing the commercial grade PEN, to be confirmed as the next step.

\section{Summary and outlook}
In the completed characterization of commercial Teonex~Q51 and Q83 grades, the DSC results indicated differences in the level of cross-linking and/or additives, while UV aging tests of Teonex~Q51 showed yellowing and a significant drop in intensity after climatic chamber exposure equivalent to 2~months of exposure to sunlight.

Then, a custom additive-free batch of PEN was successfully synthesised and matched the performance of commercial PEN foils. As the room temperature WLSE reached is still approx.~3 times lower than for TPB, examination of further steps of the synthesis is planned, including a different choice of catalyst, as well as a separate optimization of the foil pressing/stretching process.

The partial results described in this report contribute to a larger broadly-scoped R\&D program focused on WLS. Its remaining components are described elsewhere: a new table-top cryogenic facility for time-resolved WLS characterization~\cite{choudhary2024cryogenic}, and long-term stability and performance testing in a liquid argon environment~\cite{dwarf2024}.

\acknowledgments
This work has been supported from the EU’s Horizon 2020 research and innovation programme under grant agreement No 952480 (DarkWave), and from the International Research Agenda Programme AstroCeNT (MAB/2018/7) funded by the Foundation for Polish Science from the European Regional Development Fund~(ERDF). We thank Mateusz G\l owacki (Orange Polska S.A.) for aging the samples in the climatic chamber. We are grateful to Prof. Magdalena Skompska for access to the spectrophotometer, which was purchased by CNBCh (University of Warsaw) from the project co-financed by EU from the ERDF.

\bibliographystyle{JHEP}
\bibliography{biblio.bib}

\end{document}